# Generative Programming of Graphical User Interfaces

Max Schlee
Thomson Grass Valley
Brunnenweg, 9
D-64331 Weiterstadt, Germany
Phone: +49 (0) 6150 104 0
schlee2@gmx.de

Jean Vanderdonckt
Université catholique de Louvain (UCL)
School of Management (IAG)
B-1348 Louvain-la-Neuve, Belgium
Phone: +32-10/478525 – Fax: +32-10/478324
vanderdonckt@isys.ucl.ac.be

## ABSTRACT
Generative Programming (GP) is a computing paradigm allowing automatic creation of entire software families utilizing the configuration of elementary and reusable components. GP can be projected on different technologies, e.g. C++-templates, Java-Beans, Aspect-Oriented Programming (AOP), or Frame technology. This paper focuses on Frame Technology, which aids the possible implementation and completion of software components. The purpose of this paper is to introduce the GP paradigm in the area of GUI application generation. It demonstrates how automatically customized executable applications with GUI parts can be generated from an abstract specification.

## Categories and Subject Descriptors
D.2.1 [**Software Engineering**]: Requirements/Specifications – *elicitation methods (e.g., rapid prototyping, interviews, JAD)*. D.2.2 [**Software Engineering**]: Design Tools and Techniques – *user interfaces*. H.2.4 [**Database Management**]: Systems – *transaction processing*. I.3.6 [**Computer Graphics**] Methodology and Techniques – *interaction techniques*.

## General Terms: Design, Languages, Human Factors.

## Keywords: Code generation, data base, diagram transformation, feature diagram, generative programming, model-driven approach, object-oriented programming.

## 1. INTRODUCTION
Generative Programming (GP) is a software engineering paradigm based on modelling software system families. When given a particular requirement specification, it can use configuration knowledge to automatically manufacture highly customized and optimized intermediate and end products from elementary reusable implementation components [2]. It does not compete with the existing paradigms but supplements them. GP supports reusability and adjustability much better than object-oriented programming, frameworks and design patterns [5,6]. The purpose of software development automation is to speed up the development process and reduce development costs, as well as to improve software quality and error resistance. Apart from that, it helps reduce the maintenance cost and similar necessities [8]. GP represents an approach permitting the creation of whole product families. It consists of three elements (Fig. 1):

1. The left oval represents the methods used for the family member specification. It is made for users and computer experts. They use a specific language that has specific features and terms. This language is implemented as a domain-specific language (DSL). It gives the user the opportunity to describe a particular system in a most suitable way. This helps to "order" a particular system. In order to do this, a text, a form dialog, or a graphical-interactive environment can be used [5].
2. The arrow represents the configuration generator. The configuration generator automates the product assembly. It accepts a DSL specification and analyses it. Then, if necessary, it carries out a build ability check and assembles a software product from the implementation components [6].
3. The right oval describes the world of the software developer. It contains developed elementary components the system can be assembled from. They must have maximum combinability with minimum redundancy. The use of a feature diagram which graphically represents the elementary components in the form of a tree-like structure is helpful.

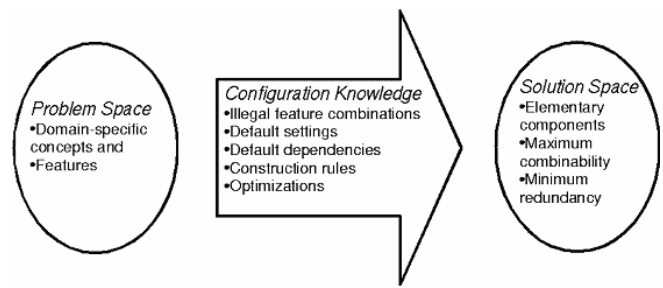

**Fig. 1. Generative domain model**

## 2. FEATURE MODELS
"*Current OO notations do not support variability modelling in an implementation independent way, e.g. when the user draws a UML class diagram, he has to decide whether to use inheritance, aggregation, class parameterization, or some other implementation mechanism to represent a given variation point*" [3]. Feature modelling is the central activity of domain engineering. It was introduced by the Feature-Oriented Domain Analysis (FODA). Feature modelling is the process of analyzing and modelling of common and variable features of concepts and their interdependencies, as well as describing their arrangement in a coherent model, the *feature model*. Feature models serve as documentation help.

With feature models, common and variable features within a system family can be modelled. The main component of the feature models are features. Apart from the name, a large amount of additional information can belong to a feature. This additional information comes mostly in the form of tables, lists, or free text. It can also be documented in diagrams or with the help of a suitable tool, for example, the feature model editor AmiEddi [11]. Feature diagrams with this additional information make up a feature model.

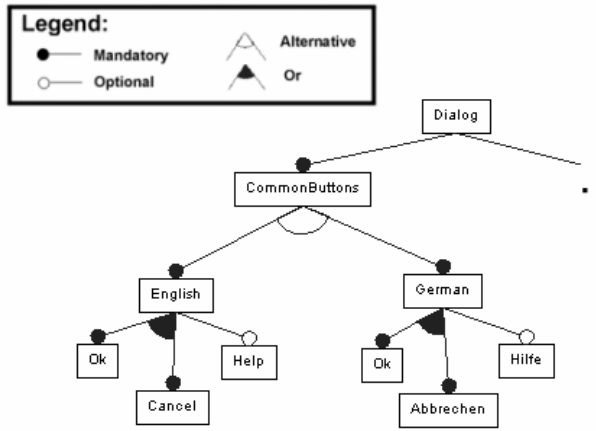

**Figure 2. Example for a feature diagram**

An example of a feature diagram is shown in a Fig. 2. It describes a part of a dialog window. Its root represents the dialog concept. The other nodes of the features are:

- *Mandatory features*: Every dialog window has the common buttons.
- *Alternative features*: A dialog window may support either English or German languages.
- *Or-features*: A dialog window may have an Ok-Button, a Cancel-Button, or both.
- *Optional feature*: A dialog window may or may not have a Help-Button

## 2.1 Dialog-based graphical-interactive DSL

The dialog-based graphical-interactive DSL permits the user to automate the whole specification development process. It is no longer necessary to see the logic or the declarations in the specifications. Moreover, the user needs no knowledge of the language used in the specification (e.g. XML). GUI elements are easier for the human perception than text specifications. When editing a text specification, it is highly possible to make mistakes, such as typing errors. Besides, the semantic rules can be violated because the user has to create and run the logic of the system that is being created in his head, which is absolutely impossible when creating complex systems. The introduction of a dialog-based graphical-interactive DSL makes it easier for the user and takes over this task. It is in charge of both dependencies appearing in the system and invalid input. The generator is becoming more and more user-friendly, the user can easily learn how to use it by trying out the available options. The user can not do anything wrong, as the system accompanies him at every step assuring an error-free creation of a specification that does not need to be verified by the generator. The generator can be used intuitively and is easy to understand. The whole process of the specification creation runs in the background, visible for the user. When creating a dialog-based graphical-interactive DSL, it is necessary to transform the feature model into GUI elements. The user does not need an advanced knowledge of the feature models used in the DSL. The transformation of the DSL must meet the following criteria:

1. The mandatory features do not appear in the dialog because they are available anyway. If such a feature does appear in the dialog, then it appears only as a group boxes title (Fig. 3).
2. The logic of the optional features is completely covered by the checkboxes. No additional logic verification is needed.
3. The Radio buttons are suitable for the group of alternative features (Fig. 4). In this case, an additional logic verification is also unnecessary because the logic coincides with the widgets.

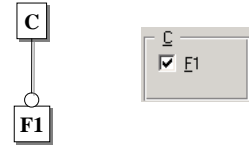

**Fig. 3. Optional feature transformation**

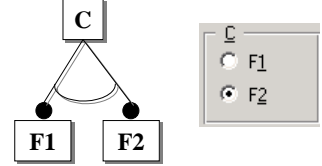

**Fig. 4. Alternate group transformation**

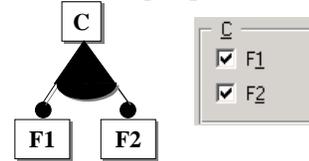

**Fig. 5. Or-group transformation**

4. A group of or-features can be represented by a check box (Fig. 5). In this case, an additional verification is necessary because the user must make sure that at least one feature is selected. Or, if the superior feature is an optional feature, the user is not allowed to deactivate it, e.g. if the user does not select one of its sub-features that are gathered in an or-group).

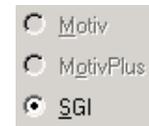

**Fig. 6. Expressing dependences with enable/disable**

5. Dependency relations can be expressed in two ways. The best way is to use enable/disable. It is impossible to select anything that is not supposed to be selected. The relations become clear when the elements are situated in one layout. The user can see how the alteration of some elements influences the others (Figs. 6, 7).

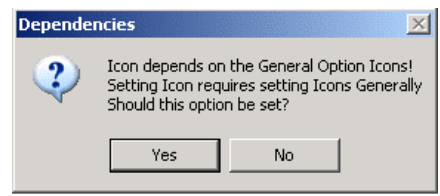

**Fig. 7. Expressing dependencies with a dialog**

Unfortunately, the enable/disable relation does not work for the elements situated in different layouts. The user cannot see what options in different layouts are influenced by the changes of an option in one layout. For this purpose, message boxes can be used. When the user selects an option, they can inform him about the consequences of his selection and offer alternatives.

## 2.2 Example of a feature diagram creation and transformation

Let's assume that the "View" Menu and all the activities that have to do with it need to be combined in one feature model and then transferred to a dialog-based graphical-interactive DSL.

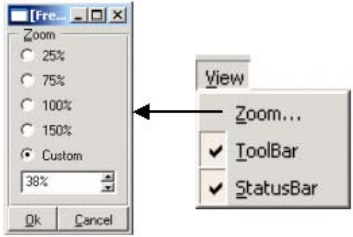

**Fig. 8. The popup menu "View"**

The popup menu "View" (Fig. 8) contains the following items:
- "Zoom" opens a dialog window from which the user can determine the size of the image on the screen.
- "Toolbar" shows or hides the toolbar.
- "StatusBar" shows or hides the status bar.

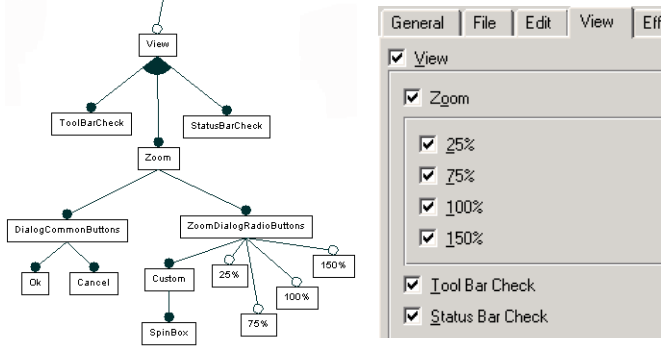

**Fig. 9. Example of a feature diagram transformation**

Fig. 9 shows the feature diagram and its related UI resulting from diagram transformation. The mandatory features are not shown in the GUI layout because it is not allowed to manipulate them. "View" is an optional feature. This is why it is represented by a check box in the layout. It is important to keep in mind that when this feature is not selected, it is impossible to select its sub-features. This is why if the "View" option is not selected, all other options of this layout become disabled. Another important thing is that the three features Tool Bar Check, Status Bar and Zoom are grouped together as an or-group. This means that at least one of them must be selected. This problem could be solved this way: if no feature from the group is selected then it signifies that these components are not available in the system. The conclusion is that View is not needed in the system. If Tool Bar Check, Status Bar and Zoom features are not selected, it is possible to deactivate the father node "View" which automatically disables sub-features. However, this state would be irreversible. The "View" component has no functional code and consists only of GUI parts, namely the menu entry where further popup menu entries for the specific components are created. If there are no such components, it means that no menu entries must be made. This is why an additional XML tag <viewmenu> is introduced in the specification. It is in charge of control over this menu entry. The optional features 25%, 75%, 100% and 150% of the "Zoom" feature are represented by check buttons. If "Zoom" feature is deactivated, these buttons are disabled.

## 3. FRAME TECHNOLOGY

Frame technology is a new generator technology that can be used with such generative paradigms as generative programming (GP), MDA and others. Frame technology deals with the concept of frames and slots. In 1974, Marvin Minsky's article "A Framework for Knowledge Presentation" [7] was published in the book "The Psychology of Computer Vision". The frame/slot approach originated in Artificial Intelligence (AI) and was then introduced to the area of picture identification. Later, it turned out that it was also possible to use this approach in the analysis and synthesis of languages [8]. A frame defines set values, the so-called *slots*. The slots of a frame can be filled with frame instances. This way, complex hierarchies can be created. The purpose of the frames is to classify the scene descriptions or texts based on their patterns [5]. This concept is very powerful for the representation of texts. Frame technology can be used while working with the generative paradigm [5]. This technology proved effective and showed decent results in industrial use [1]. Frame technology is well suited for generative programming. Feature models [2] possess all necessary information to build a frame hierarchy.

## 4. ANGIE-BASED GUI-GENERATOR (ABA)

In ABA, the generative domain model is divided so that the problem area is projected on the GUI of the specificator that is connected to the GUI generator. Most of the configuration knowledge such as the default settings, dependent features, illegal feature combinations as well as optimizations is taken over by the specificator. The construction rules are carried out with the help of ANGIE script functions. The solution area is projected on ANGIE frames. The frame contents include such files as C++, make files, XPM and Developer Studio workspace. Make files have two versions one for Windows and the other for Linux. The Microsoft Developer Studio Workspace Files (.DSW and .DSP) are created to make further work on the prototype easier (Fig. 10).

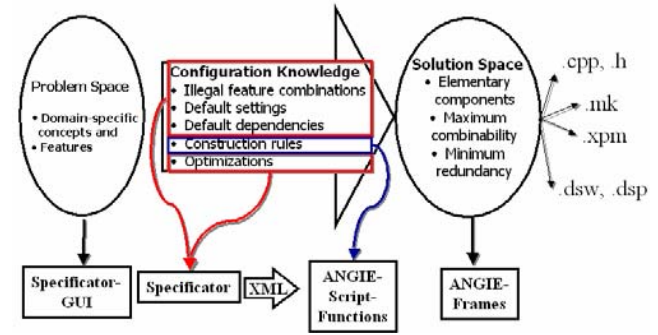

**Fig. 10. The distribution of generative domain model in ABA**

For smooth and productive work with ABA, the following software support is required: Windows 9.x, NT, 2k, XP, ANGIE R2.1, Qt 2.3.0, Visual C++ 6.0. In order to analyze GP, a tool named ANGIE was used (developed by the Delta Software Technology GmbH) [4]. The generated GUI prototypes are completely based on Qt and can be compiled both in Windows and in Linux. This ability of Qt to function with different platforms was the only reason why this framework was chosen for this project. Theoretically, any library and programming language could be used instead of Qt and C++.

## 4.1 The generation process in ABA

The whole development process runs in the background and is visible for the user. This is the reason why the generator can be used intuitively and is easily understandable. The base of the process is the ABA user interface. It controls the whole generation process. One of the most important tasks of the ABA user interface is the creation of the XML specification that is then transferred to the ANGIE-part of the generator. There, the system creates the source code according to the selected specification. If the user wishes, the "Make" process is also carried out. When it is completed, the ABA UI can start the created application (Fig. 11).

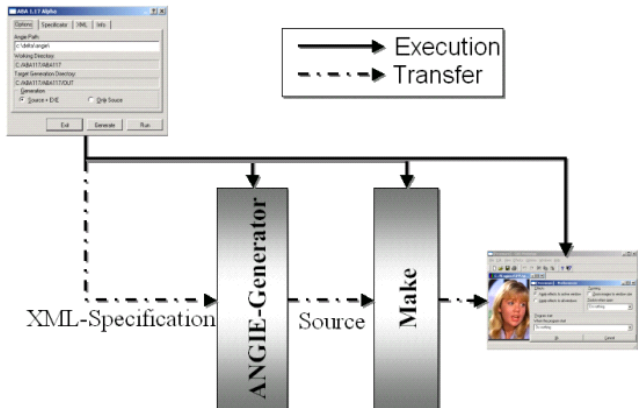

**Fig. 11. The generation process in ABA**

## 4.2 The Process of Creating a XML-Specification

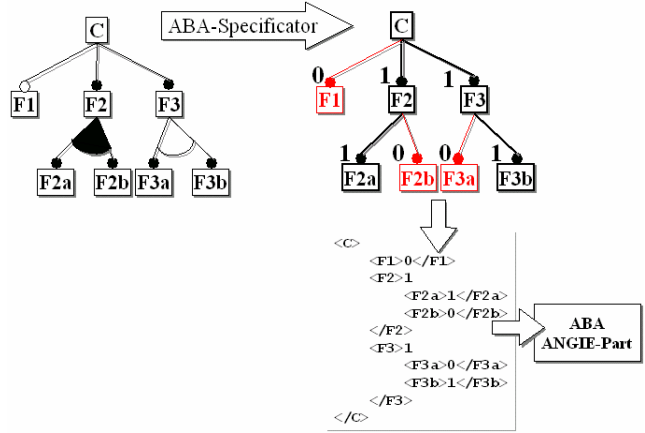

**Fig. 12. The creation process of the XML-specification**

The main purpose of the specificator is to create a specification of a system. In order to accomplish this, it is necessary to remove all the variable parts from the feature diagram. This process is called specialization. The first part of the figure shows an example of a feature model that undergoes a specialization process with the help of the ABA specificator (Fig. 12). A tree-like structure appears. Its "leaves" are "weighed down" by zeros and ones. If a feature is annotated with 0, then the generator gets the message that the component associated with this feature can not be in the system. Alternately, the features marked with 1 are required by the system. This tree-like structure is described with XML. Based on the XML presentation, the user can see the whole structure of the system that is being generated and can imagine the components the system consists of. Based on this presentation, the user sees the non-existing components and the corresponding parts of the system where these components will not be situated.

## 5. CONCLUSION

The feature diagram of the entire system of possible GUI prototypes consists of over 200 features. The generator permits the creation of $V \approx 5 * 10^{17}$ prototype variants, better than $V \approx 6 * 10^7$ prototype variants which can be created with Microsoft AppWizard. The frame borders are labeled in the comments, in the source of the generated prototypes. The frames can be updated in the generator after the editing of source. This makes Roundtrip-Engineering possible and allows to expand the generator when necessary. During the MiniABA-Project [6], we successfully tested the option of generating GUI as Resources.

## ACKNOWLEDGEMENTS

This research was carried out by Max Schlee as part of the PoLITe project, whose members are gratefully acknowledged. PoLITe stands for "The Development and Testing of the Manual for the Software Product Line Implementation Technologies". This project was supported by the Foundation for Innovation in Rhineland-Palatinate, Germany. The development of the software lines is documented at the Fraunhofer Institute for the Experimental Software Engineering (IESE) and at the University of Applied Sciences in Kaiserslatern (its branch in Zweibrücken).